\documentclass[aps,prb,twocolumn,amsfonts,amsmath,amssymb]{revtex4}

\usepackage{graphicx}
\usepackage{hyperref}

\begin{document}
\title{The construction of a reliable potential for  GeO$_2$ from first-principles}
\author {D. Marrocchelli $^1$, M. Salanne$^{2,3}$, P.A. Madden $^4$, C. Simon$^{2,3}$ and P. Turq$^{2,3}$}

\affiliation{$^1$ School of Chemistry, University of Edinburgh, Edinburgh EH9 3JJ, UK}
\affiliation{$^2$ UPMC Univ Paris 06, UMR 7612, LI2C, F-75005, Paris, France}
\affiliation{$^3$ CNRS, UMR 7612, LI2C, F-75005, Paris, France}
\affiliation{$^4$ Department of Materials, University of Oxford, Parks Road, Oxford OX1 3PH, UK}

\begin{abstract}
The construction of a reliable potential for GeO$_2$, from first-principles, is described. The obtained potential, which includes dipole polarization effects, is able to reproduce all the studied properties (structural, dynamical and vibrational) to a high degree of precision with
a single set of parameters. In particular, the infrared spectrum was obtained with the expression proposed for the dielectric function of polarizable ionic solutions by Weis {\it et al.} [J.M.~Caillol, D.~Levesque and J.J.~Weis, {\it J. Chem. Phys.}, {\bf 91}, 5544 (1989)]. The agreement with the experimental spectrum is very good, with three main bands that are associated to tetrahedral modes of the GeO$_2$ network. Finally, we give a comparison with a simpler pair-additive potential.

\end{abstract}

\maketitle

\section{Introduction}

In  the vitreous and liquid states, at ambient pressure, germania (GeO$_2$) (a
close structural analog of silica (SiO$_2$)) forms a tetrahedrally coordinated
three-dimensional network \cite{micoulaut2006b}. Because of its lower abundance,
its usage in practical applications is much less widespread than for the SiO$_2$.
Still, GeO$_2$ is used in several fields, mainly related to optical technologies.
For example, a mixture of SiO$_2$ and GeO$_2$ allows precise control of refractive
index in optical fibres and waveguides. It is of interest to develop simulation
methods to allow a detailed examination of the local structure in such mixtures and
to predict the infrared spectrum, since this determines the long wavelength limit
for their use as optical fibres.

The essential similarity between the structures of glassy GeO$_2$, SiO$_2$ and
BeF$_2$ was made clear by the study of vibrational properties by Galeener and
co-workers \cite{galeener1983a} who showed that the inelastic neutron, infrared and
Raman spectra of the different materials were closely related. More recent infrared
studies of GeO$_2$ have also been reported \cite{kamitsos1996a,teredesai2005a}.
Recently a full set of partial structure factors were determined in this system by
using the method of isotopic substitution in neutron diffraction experiments
\cite{salmon2006b,salmon2007a}. The results show that the tetrahedral network
structure is based on corner sharing Ge(O$_{1/2}$)$_4$ tetrahedra with a Ge-O
average distance of 1.73~\AA\ and with a mean inter-tetrahedral Ge-O-Ge angle of
132$^\circ$. They also show that the topological and chemical ordering in the
network display two characteristic length scales at distances greater than the
nearest neighbour.

Several  molecular dynamics (MD) simulations have been undertaken to study the
structural and vibrational properties of the disordered phases of GeO$_2$.
Classical MD simulations were performed on glassy and liquid GeO$_2$
\cite{micoulaut2006a,peralta2008a}, using pairwise additive potentials with partial
charges developed by Oeffner and Elliott (OE) for modeling the $\alpha$-quartz and
rutilelike phases of GeO$_2$ \cite{oeffner1998a}. More recently, first-principles
molecular dynamics (FPMD) of glassy GeO$_2$ have also been reported
\cite{giacomazzi2005a,giacomazzi2006a}. In principle, the amount of empirical
information needed to set up a first-principles calculation is minimal and it would
normally be the method of choice to study the physico-chemical properties of
condensed phase systems. However, it is computationally very expensive compared to
classical molecular dynamics, which is a major drawback when dealing with glassy
systems. Classical and FPMD simulations were also compared one with each other in a
combined study\cite{hawlitzky2008a} for temperatures above 2530~K. In this study
{\it apparent} inconsistencies between the properties predicted by the classical
simulations of different authors are noted. In reality, two {\it different}
parameter sets were proposed by OE: an original one, which was fitted from an {\it
ab initio} energy surface, and a so-called rescaled one, which was developed from
the previous one in order to give a better reproduction of the vibrational
properties. For example, the partial charge of the germanium ion was shifted from
1.5~$e$ in the original potential to 0.94174~$e$ in the rescaled potential. The
apparent inconsistencies arise because different classical potentials were being
used. Our objective is a single potential which will give a good description of
structure and dynamics and allow long simulations on systems of many atoms.

The structure of the simulated system can  directly be compared with the neutron /
X-ray diffraction experiments. Both classical and FPMD simulation provide structure
factors which are in rather good agreement with experiments. For the classical
molecular dynamics, the most noticeable difference is a shift to high $q$-values of
the first sharp diffraction peak (FSDP), which was observed with both versions of
the OE potential \cite{micoulaut2006a,peralta2008a}. As this peak is associated to
medium range distances \cite{salmon2006b}, this means that the Ge-O-Ge bond angle
is too large and the topology of the tetrahedral network is not correctly
described. The FPMD simulations provide a diffraction pattern with improved
agreement \cite{giacomazzi2006a}, but the small size / timescales explored lead to
large statistical uncertainties, mainly in the region of this FSDP. The vibrational
properties can also be straightforwardly calculated. The vibrational density of
states (VDOS) was estimated from FPMD and classical MD using the rescaled OE
potential, with an overall reasonable agreement with the available experimental
data extracted from inelastic neutron scattering experiments. The infrared
absorption spectrum, which originates from the polarization fluctuations associated
with motion of the atomic charges, has only been reported from the FPMD simulation.
The agreement with experiment \cite{galeener1983a} was overall good, despite a
small shift of the peaks towards lower frequencies.

The dynamical properties which involve structural rearrangements are out of reach
of the FPMD. The only studies on GeO$_2$  employ classical MD,  and were aimed at
determining the diffusion coefficients in the liquid state
\cite{micoulaut2006a,hawlitzky2008a}. Hawlitzky {\it et al.}\cite{hawlitzky2008a}
observed important differences between their values and those obtained by Micoulaut
{\it et al.} \cite{micoulaut2006a};
 their hypothesis for rationalizing this discrepancy was inadequate equilibration in the latter. In fact, as noted above, it is due to the use of the two different OE potentials in
these studies. Indeed, in simulations performed with the rescaled OE potential the
fluidity of the system is enhanced by more than one order of magnitude compared to
the results obtained with the original OE potential.

In this work we describe the parameterization of a new classical MD interaction
potential for GeO$_2$, by using some techniques we have developed for halides and
oxides \cite{madden2006a,jahn2007b}, which is entirely based on first-principles
electronic structure calculations. In order to provide an accurate, transferable
description of the interactions, these potentials include dipole polarization
effects and the ions carry full valence charges. The various structural, dynamical
and vibrational properties are then compared with the corresponding experimental
results. In future work we will use such potentials to study mixtures and the
effect of pressure on glassy GeO$_2$

\section{Interaction potential development}
\label{Section II} The interaction model used in this work is related to that used
in previous work\cite{castiglione1999a,castiglione2001a}, with a couple of minor
differences. The model (known as DIPPIM) includes a pair potential, together with
an account of the polarization effects which result from the induction of dipoles
on the O$^{2-}$ ions. The parameters for these potentials were obtained by
force-matching them to first-principles reference data \cite{madden2006a}. 
Such an approach was successfully applied in the case of other oxide materials \cite{jahn2007b}.
Tangney and Scandolo \cite {tangney2002a} also used a similar interaction model to study silica, 
in which all the parameters including partial charges on the ions were fitted against first-principles data. 
Here we use formal ionic charges (Ge$^{4+}$, O$^{2-}$) which should ensure a better transferability.
In the next paragraphs we will give a brief description of the model used, the
first-principles reference calculations and the force-matching procedure.

\subsection {The DIPole-Polarizable Ion Model (DIPPIM)}
The interatomic potential is constructed from four components: charge-charge,
dispersion, overlap repulsion and polarization. The first three components are
purely pairwise additive:
\begin{equation}
V^{\rm qq}=\sum_{i\leq j}\frac{q_i q_j}{r_{ij}}
\end{equation}
where $q_i$ is the {\it formal} charge on ion $i$. The dispersion  interactions
include dipole-dipole and dipole-quadrupole terms
\begin{equation}
V^{\rm disp}=-\sum_{i\leq j }[\frac {f_6^{ij} (r^{ij})  C_6^{ij}}{r_{ij}^6
}+\frac{f_8^{ij} (r^{ij} ) C_8^{ij}}{r_{ij}^8 }].
\end{equation}
Here $C_6^{ij}$ and $C_8^{ij}$ are the dipole-dipole and  dipole-quadrupole
dispersion coefficients, respectively, and the $f_n^{ij}$ are the Tang-Tonnies
dispersion damping function, which describe short-range corrections to the
asymptotic dispersion term. The short range repulsive terms are approximately
exponential in the region of physical interionic separations. The full expression
used here for the short range repulsion is:
\begin{eqnarray}
  V^{\rm rep}= \sum_{i\leq j} \frac {A^{ij} e^{-a^{ij} r_{ij}}} {r_{ij}}+ \sum_{i\leq j}B^{ij} e^{-b^{ij} r_{ij}^2 },
\end{eqnarray}
where the second term is a Gaussian which acts as a steep  repulsive wall and
accounts for the anion hard core; these extra terms are used in cases where the
ions are strongly polarized to avoid instability problems at very small anion-cation separations \cite{castiglione1999a}. The polarization part of the
potential incorporates dipolar effects only. This reads:
\begin{eqnarray}
V^{\rm pol}&=&\sum_{i,j}-\left( q_i\mu_{j,\alpha} f_4^{ij}(r_{ij})- q_j\mu_{i,\alpha} f_4^{ji}(r_{ij})\right)T_{\alpha}^{(1)}({\bf r}_{ij}) \nonumber \\
& &-\sum_{i,j}\mu_{i,\alpha} \mu_{j,\beta} T_{\alpha\beta}^{(2)}({\bf
r}_{ij})+\sum_i \frac{1}{2\alpha_i}\mid {\boldsymbol \mu}_i \mid^2.
\end{eqnarray}
Here $\alpha_i$ is the polarizability of ion $i$, ${\boldsymbol \mu}_i$ are the dipoles and ${\bf T}^{(1)}$, ${\bf T}^{(2)}$ are the charge-dipole and dipole-dipole interaction tensors: \\
\begin{equation} T_{\alpha}^{(1)}({\bf r})=-r_{\alpha}/r^3 \;\;\;\;\;\;\;\;\; T_{\alpha \beta}^{(2)}({\bf r})=(3r_{\alpha} r_{\beta}-r^2\delta_{\alpha \beta})/r^5. \end{equation}
The instantaneous values of these moments are obtained by  minimization of this
expression with respect to the dipoles of all ions at each MD timestep. This  ensures that we regain the condition that the dipole 
induced by an electrical field ${\bf E}$ is $\alpha {\bf E}$ and that the dipole values are mutually consistent.
The short-range induction effects on the dipoles
are taken into account by the Tang-Toennies damping functions:
\begin{equation}
 f_n^{ij}(r_{ij} )=1-c^{ij} e^{-d^{ij} r_{ij}} \sum_{k=0}^n \frac{(d^{ij}
 r_{ij})^k}{k!}.
\end{equation}
The parameters $d^{ij}$ determine the range at which the overlap of the charge
densities affects the induced dipoles, the parameters $c^{ij}$ determine the
strength of the ion response to this effect. It is important to notice that
anion-anion damping terms have been taken into account, contrary to what was done
in \cite{castiglione1999a}. The addition of anion-anion damping terms was found to
greatly improve the ability to match the first-principles data.

\subsection{The DFT reference calculations}
The parameters in the interaction potential are determined by matching the dipoles
and forces on the ions calculated from first-principles on condensed phase ionic
configurations \cite{madden2006a,jahn2007b}. Starting from the empirical pair
potential in \cite{micoulaut2006a}, we obtained atomic configurations for GeO$_2$
by running short MD simulations on small cells (150 ions); a total of two liquid
configurations were obtained for GeO$_2$; for each of these, the Hellman-Feynman
forces acting on individual ions of the simulation cell were calculated using
planewave-DFT code CPMD \cite{CPMD}. In all the calculations we used
norm-conserving pseudopotentials and planewave energy cut-offs of $1360\;eV$; all
calculations were performed using the generalized gradient approximation (GGA)
according to the Perdew, Burke and Ernzernhof (PBE) scheme. For the calculation of
first-principles dipoles, the Kohn-Sham orbitals are localized via a Wannier
transformation to construct maximally localized Wannier functions (MLWF) and the
dipoles determined from the positions of the centres of the Wannier functions
associated with each ion \cite{aguado2003b}.

\subsection{Fitting procedure}
\begin{figure}[htbp]
\begin{center}
\includegraphics[width=9cm]{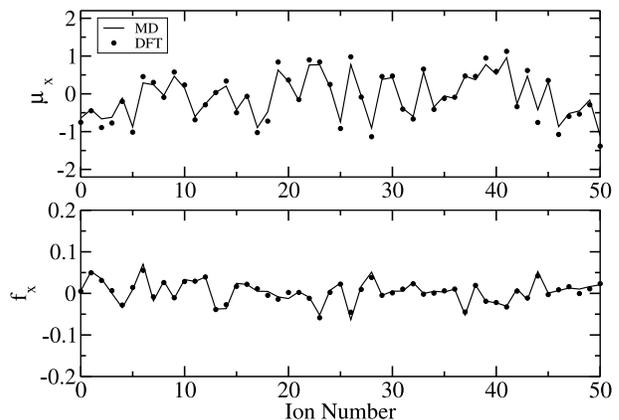}
\end{center}
\caption{ Agreement between the DFT calculated  dipoles (top panel) and
forces (bottom panel) and those predicted by a fitted DIPPIM potential model for a
set of 50 representative ions. In each panel the ordinate gives the ion number and
the abscissa gives the DFT values of the x component of the forces and dipoles as
points and the fitted values as the continuous line.} \label{FigureII}
\end{figure}

The potential parameters   are optimized by fitting the forces and dipoles, obtained
 with the DIPPIM potentials for the reference configurations, to the respective
  results from the DFT calculations; the 2 configurations provide a total of about
   1800 data points to fit, comprising three Cartesian force components of each individual
 ion and three components for the dipoles. While most of the potential
parameters are left as free parameters in the fits, there are some exceptions.
The O$^{2-}$ polarizability, for instance, was fixed to $\alpha_{{\rm O}^{2-}}=11\;a.u.$,
 i.e. the value Salmon $et\; al$ obtained experimentally in \cite{salmon2006b}.
 This value for the oxide polarizability is also compatible with the range of
  values obtained for this quantity in several magmatic melts from first-principles calculations \cite{salanne2008e}.
One problem in DFT calculations is the uncontrolled representation of the
dispersion interaction. Although dispersion only contributes a tiny fraction to the
total energy, it has a considerable influence on phase transition pressures and on
the material density and stress tensor. For this reason, we decided not to include
the dispersion parameters in the fits as discussed in \cite{madden2006a} and to add
them afterwards. We used the parameters in \cite{norberg2009a} for the oxygen terms
and rescaled these by the Ge polarizability for the anion-cation terms. The
Gaussian parameters too were added after the fit and then the comparison with the
first-principles data was run again to check that its quality remained unchanged.
\newline
In Figure \ref{FigureII} we  report the agreement between the DFT calculated forces
and dipoles and those predicted by the fitted potential, for a set of
representative ions; the abscissa shows the $x$ component of the forces and
dipoles, while the ordinate is simply the ion number. The quality of the
representation is quite good and comparable with the one obtained in
\cite{wilson2004a}; this can be regarded as a very good result, considering the
simplicity of the model (both ion shape deformation effects and quadrupoles are not
taken into account in the present model \cite{madden2006a,jahn2007b}). The
parameters obtained for the DIPPIM potential are summarized in Table \ref{TableI}.

\begin{table}
\caption{\label{TableI} Parameters  in the repulsive, polarization and dispersive
parts of the DIPPIM potential. All values are in atomic units. The oxide ion
polarizability was set to 11 au.}
\begin{tabular}{@{}cccc}
\hline
         & O-O & O-Ge & Ge-Ge      \\
\hline
$A^{ij}$ &   17.78   & 40.89 & 2958   \\
$a^{ij}$ &   3.893   & 1.186 & 5.96     \\
$B^{ij}$ & 50000 & 50000 & 0 \\
$b^{ij}$ & 1.1   & 1.75   &   /     \\
\\
$C_6^{ij}$      & 44  & 4  & 0      \\
$C_8^{ij}$      & 853 & 50 & 0   \\
$c_{\rm 6}^{ij}$ & 1.0 & 1.0 & /    \\
$c_{\rm 8}^{ij}$ & 1.0 & 1.0 & /    \\
$d_{\rm 6}^{ij}$ & 1.0 & 1.5 & /    \\
$d_{\rm 8}^{ij}$ & 1.0 & 1.5 & /    \\
\\
$d_{\rm pol}^{ij}$ &  2.208 & 1.977 & /    \\
$c_{\rm pol}^{ij}$ &  1.770 & 1.709 & /  \\
\hline
\end{tabular}
\end{table}

\section{Simulation details}

We performed MD simulations in the  $NVT$ ensemble through the use of the
Nos\'e-Hoover chain thermostat method \cite{nose1984a,martyna1992a}. The simulation
cell contained 288 oxide and 144 germanium ions, and its volume was set to
6856~\AA$^3$ in order to match the experimental density at 300~K, which is
3.66~g.cm$^{-3}$. We used a time step of 1~fs to integrate the equations of motion
and the minimization of the polarization energy was carried out with a conjugate
gradient method. The system was studied in both the liquid and amorphous phases.

For the amorphous phase, the system  was first equilibrated for 2~ns at the
temperature of 4500~K. We have then cooled down the system with a cooling rate of
3.33$\times$10$^{11}$~K.s$^{-1}$, by rescaling the velocities and decreasing the
target temperature of the thermostat in order to reduce the temperature by 50~K
every 150000 MD steps. Finally, a 5~ns long simulation was undertaken at 300~K,
from which we computed all the data presented in this article. This procedure was
performed with two different potentials, the DIPPIM, and the original OE, which has
extensively been used in the literature \cite{oeffner1998a}. A simulation at 300~K
was also performed with the rescaled OE potential (with one of the configurations
obtained from simulations with the original OE potential as a starting
configuration) for the study of the vibrational properties.

For the liquid, a series of long  simulations with the DIPPIM potential at the
temperatures of 3600~K, 3800~K, 4000~K, 4200~K, 4400~K, 4500~K, 4600~K, 4800~K and
5000~K was undertaken. In all the cases, an equilibration run was performed so that
the slower species, Ge, moved on average, at least a distance of $5.5$\AA; a
subsequent run of the same length was made to accumulate enough statistics for the
mean-squared displacements curves. At these temperatures, the runs were between
100,000 and 1,000,000 steps. Both the original and rescaled OE potentials were also
used within the same simulation conditions at a temperature of 3600~K.

\section{Static structure}

Neutron diffraction experiments with the isotope substitution technique allowed
Salmon to determine very precisely the full set of partial radial distribution
functions (RDF) \cite{salmon2006b}. These provide a good check of the validity of
the potential.

\begin{figure}[htbp]
\begin{center}
\includegraphics[width=9cm]{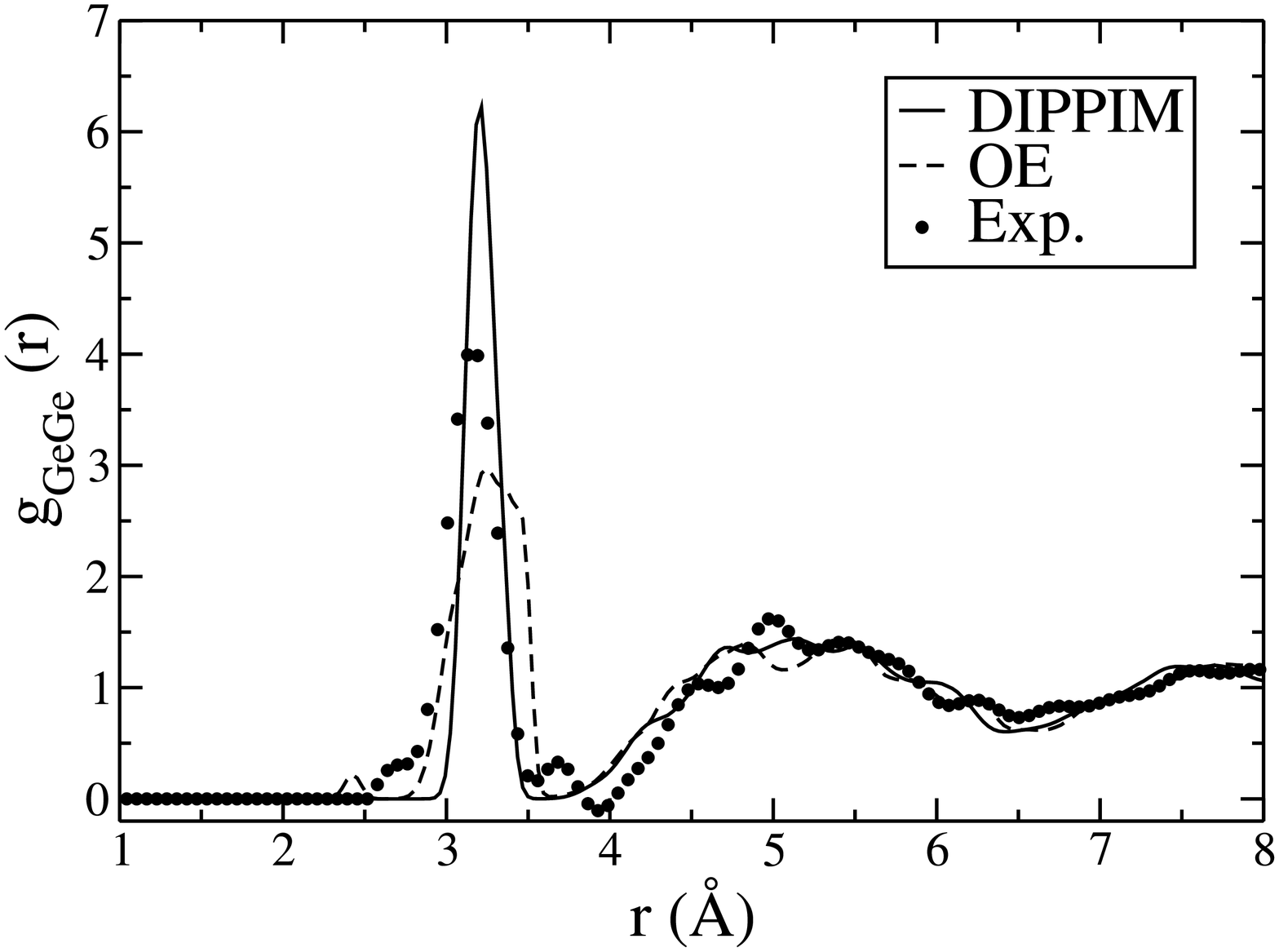}
\includegraphics[width=9cm]{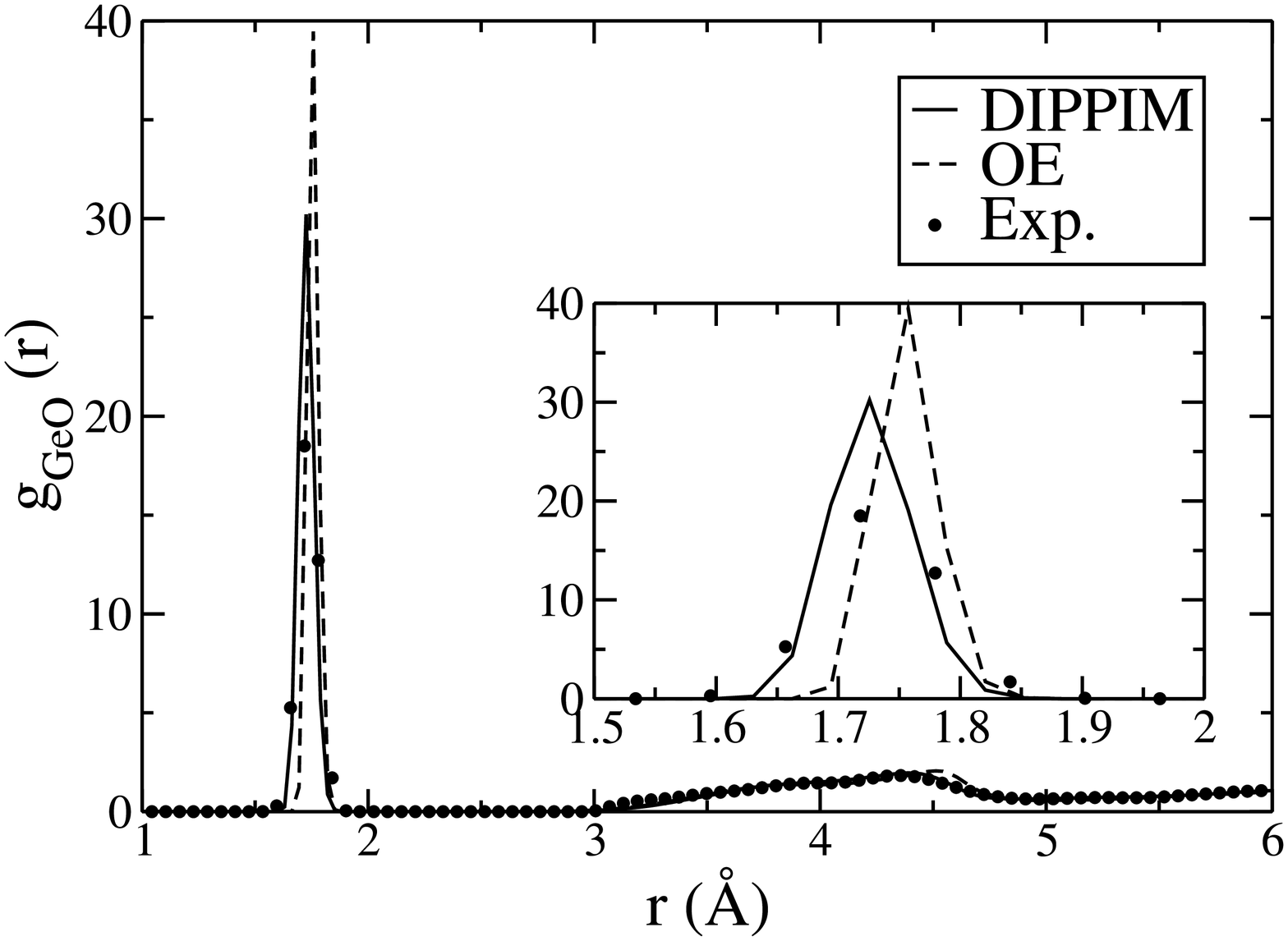}
\includegraphics[width=9cm]{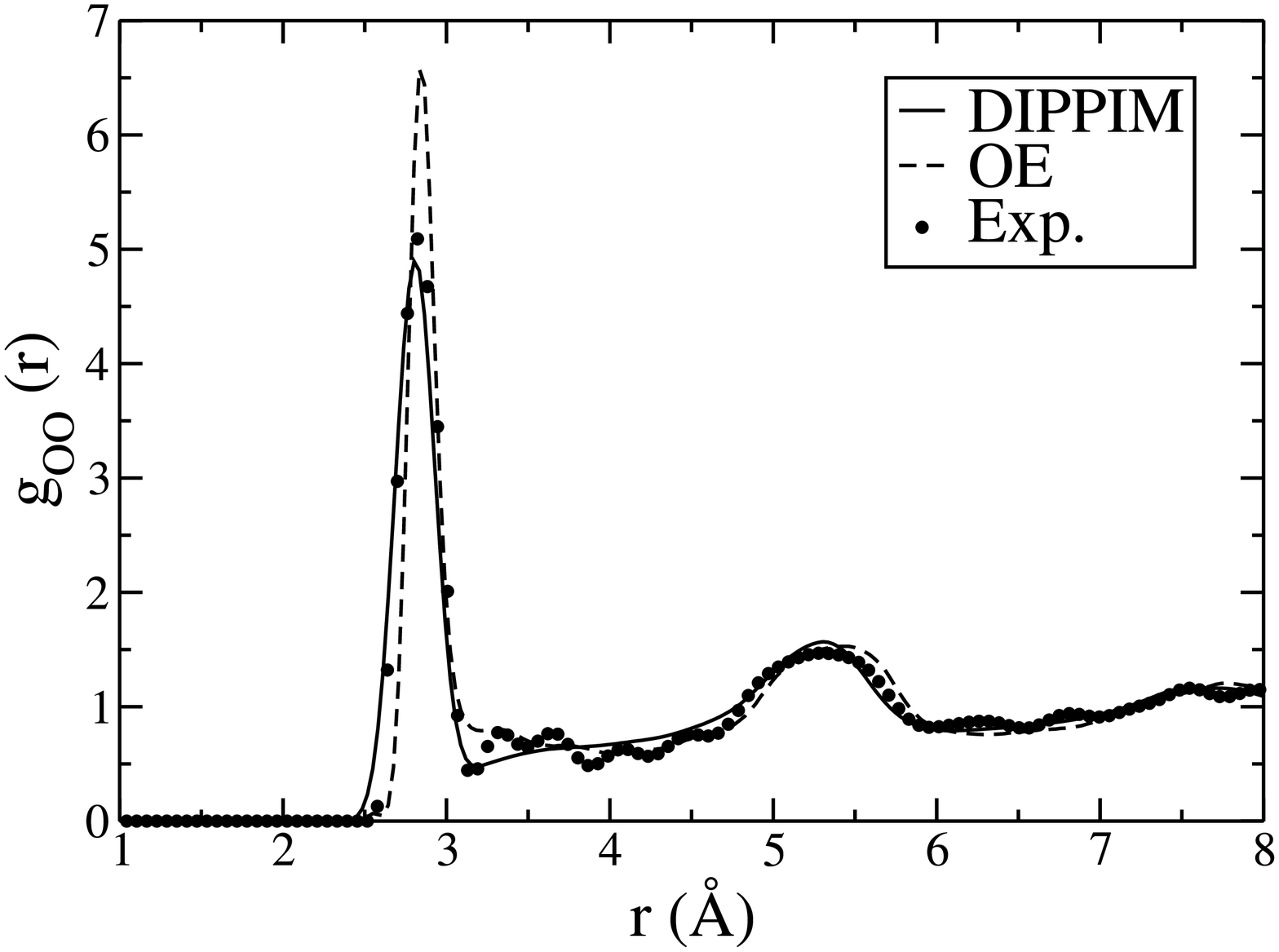}
\end{center}
\caption{ Radial distribution  functions obtained from the 300~K
simulations with the DIPPIM and original OE interaction potentials. These are
compared to the experimental ones obtained from neutron diffraction experiments
\cite{salmon2006b}.} \label{rdf}
\end{figure}

The experimental RDFs are given in figure  \ref{rdf}, together with the ones
obtained from our simulations at 300~K involving the DIPPIM and the original OE
potential; it can be readily observed that our DIPPIM potential gives a closer
match with the experimental data than the OE potential, for which all the
characteristic first-neighbour distances are slightly overestimated. Concerning the
shape of the function, the peaks obtained for the Ge-Ge and Ge-O partials are too
sharp but it has be be remembered that other factors, such as system size and
cooling rate, also play a role in this aspect of the comparison. The O-O partial
RDF is almost in perfect agreement with the experimental one, this is very
important since the O$^{2-}$ anions constitute 2/3 of the atoms of the system and
their packing arrangement determines the arrangement of the tetrahedral network.
The OE potential gives a reasonable description of the static structure (though
significantly poorer than the DIPPIM one); one might therefore be tempted to use
this potential, due to its simpler form and faster computational time; we will see
in the next sections that a reasonable reproduction of the structure might not be
sufficient for it to be capable of predicting the dynamical and vibrational
properties.

\section{Dynamics}
\begin{figure}[htbp]
\begin{center}
\includegraphics[width=9cm]{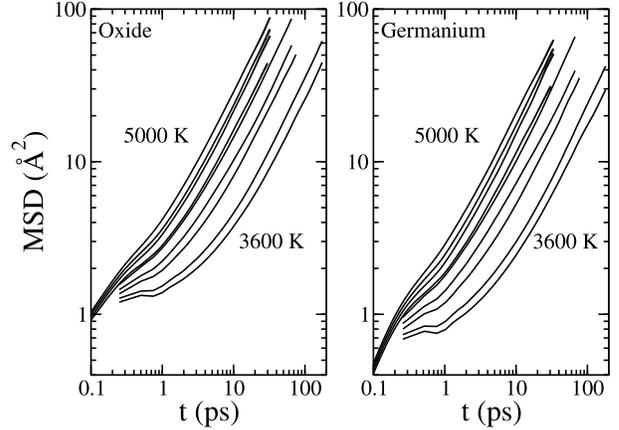}
\end{center}
\caption{ Time dependence of the  mean squared displacement curves for
both the oxygen and germanium ions. The temperatures ranges between 3600 K and 5000
K. } \label{msd}
\end{figure}

In this section we will calculate diffusion coefficients for the liquid and compare
them both with those obtained with the OE potentials and with the experimental
values. The diffusion coefficient of species $\alpha$ can be obtained from the
slope of the mean squared displacement at long times, i.e.
\begin{equation} D_\alpha=\lim_{t\rightarrow \infty} \frac{1}{6t}  \left <r_{\alpha}^2(t) \right >
, \label{eq-diffusion} \end{equation}
where
\begin{equation}  \left <r_{\alpha}^2(t) \right > = \frac{1}{N_{\alpha}}  \sum_{i=1}^{N_{\alpha}}  \left < | r_i(t)-r_i(0)^2 \right | > . \end{equation}
The mean squared displacement curves were  evaluated from a series of long
simulations at temperatures between 3600 K and  5000 K; these are shown on a
log-log scale on figure \ref{msd}. No plateau is observed even at the lowest
temperature, which shows that the system remains liquid in the range of the study.

\begin{figure}[htbp]
\begin{center}
\includegraphics[width=9cm]{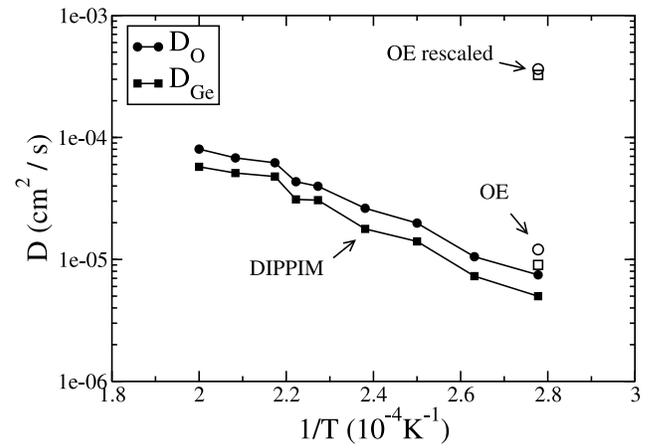}
\end{center}
\caption{ Diffusion coefficients obtained with a DIPPIM potential at different temperatures; data points obtained with both the original and rescaled OE potentials at 3600~K are also presented. }
\label{diff-coeff}
\end{figure}

The diffusion coefficients were extracted  using equation \ref{eq-diffusion}; these
are reported in figure \ref{diff-coeff} and compared with those obtained with the
OE and rescaled OE potentials. Figure \ref{diff-coeff} shows many interesting
features: first of all, our DIPPIM potential gives approximately the same value for
the diffusion coefficients as the original  OE potential. Hawlitzky
$et\;al$\cite{hawlitzky2008a}  used this potential to calculate the
 diffusion coefficients and showed that the obtained values are in good agreement
 with the experimental ones obtained by  converting viscosity data;
  we can therefore conclude that our DIPPIM potential gives a good agreement with the experimental data as well.
   A direct comparison  with experimental data might be done by using the
    only published datum ($D_{\rm O} = 7\times 10^{-10}$~cm$^2$s$^{-1}$)
     on the oxygen diffusion constant \cite{tokuda1963a}, at $T$=1440~K. However, such a low diffusion coefficient would involve
     an impossibly long simulation run. \\
It is also interesting to see that the rescaled OE potential yields too fluid a
melt if compared with the original OE and DIPPIM potentials and with experimental
data. This potential was used by Micoulaut {\it et al.}\cite{micoulaut2006a} and
they indeed obtained diffusion coefficients which were more than one order of
magnitude larger than the values obtained by Hawlitzky {\it et
al.}\cite{hawlitzky2008a} (who incorrectly explained this discrepancy in terms of
lack of equilibration of the simulations, it is obvious here that the real reason
was the difference between the two potentials). The so-called rescaled potential
was obtained by taking the normal OE potential and rescaling the ionic charges, the
van der Waals coefficient and the repulsive parameters in order to get a better
agreement with the experimental vibrational spectrum for the crystal. To do this
the parameters which are responsible for the strength of the the O-O and O-Ge short
range repulsion were lowered by a factor of approximately 2.5; as a consequence,
these ions can diffuse much more easily in a softer environment.

\section{Infrared spectrum}
The infrared spectra of ionic melts with polyvalent cations exhibit discrete bands
attributable to the vibrational motion of the local coordination complexes around
the cations. They originate from the polarization fluctuations associated with
motion of the ionic charges. The inclusion of polarization effects for the oxide
ions  in our model may influence the predicted spectrum in two ways
\cite{wilson1996a}. First, the interactions of the oxide ion dipoles may alter the
local structure of the network and the strength of the bonds, which may introduce a
shift of the vibrational frequencies. Second, the induced dipoles will themselves
be responsible for absorption, as they too contribute to the total polarization
fluctuations.
\begin{figure}[htbp]
\begin{center}
\includegraphics[width=9cm]{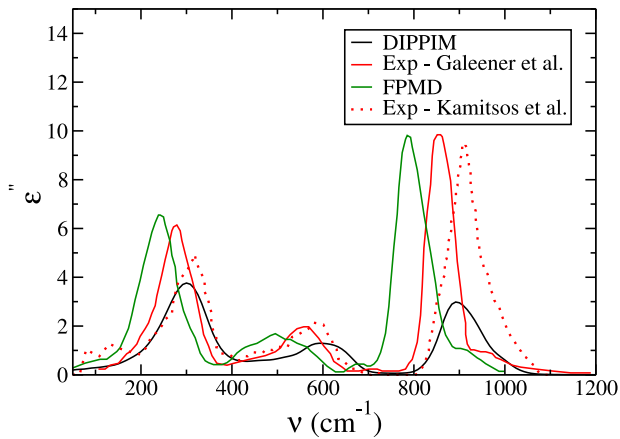}
\end{center}
\caption{ Imaginary part of the dielectric function, calculated for simulations carried out with the DIPPIM potential, compared to FPMD simulation \cite{giacomazzi2005a} and experimental results \cite{galeener1983a,kamitsos1996a}.}
\label{dielectric}
\end{figure}

The absorption coefficient in the presence of these extra moments is calculated
from the imaginary part of the total dielectric function $[n(\nu)\alpha(\nu)=
2\pi \nu \Im(\epsilon(\nu))]$, which can be determined following Caillol, Levesque
and Weis\cite{caillol1989a,caillol1989b} as
\begin{eqnarray}
\epsilon(\nu)-\epsilon_\infty&=&\frac{\beta}{3 \epsilon_0 V}\left(\langle {\bf M}(0)^2\rangle +2\pi \imath \nu \langle {\bf M}\cdot{\bf M}\rangle_\nu  \right.  \\
&&\left.+2\langle {\bf M}\cdot{\bf J}\rangle_\nu+\frac{\imath}{2\pi\nu}\langle {\bf J}\cdot{\bf J}\rangle_\nu \right) \nonumber
\end{eqnarray}

where
\begin{equation}
\langle {\bf J}\cdot{\bf J}\rangle_\nu = \int_0^\infty e^{2\pi\imath\nu  t} \langle {\bf J}(t)\cdot {\bf J}(0)\rangle {\rm d}t,
\end{equation}
${\bf J}(t)$ is the charge current ${\bf J}(t)=\sum_{i=1}^N q_i{\bf v}_i(t)$ and ${\bf M}(t)$ is the total system induced dipole moment, ${\bf M}(t)=\sum_{i=1}^N{\boldsymbol \mu}_i(t)$.

Figure \ref{dielectric} shows the imaginary part of the total dielectric function calculated  for simulations
carried out with the DIPPIM potential, compared to the one obtained from FPMD simulations \cite{giacomazzi2005a,giacomazzi2006a} and from several experiments \cite{galeener1983a,kamitsos1996a} (the spectrum proposed by Kamitsos {\it et al.} shows the absorption coefficient, here we transformed it into the dielectric function to facilitate comparisons). Three main bands are observed on the
spectra, in good agreement with the other studies. The corresponding
characteristic peak frequencies for the absorption spectra are summarized in table \ref{TableII}. The
agreement is very satisfactory; the frequencies obtained match very closely to the
various experimental results (even though these show a significant scatter), while
the FPMD simulation seems to systematically underestimate them. The relative
intensities of the bands is also in good agreement with experiment, though the
highest frequency band at around 894~cm$^{-1}$ is not intense enough in our
spectrum.

\begin{table}
\caption{\label{TableII} Characteristic frequencies of  the imaginary part of the dielectric function as
obtained from the DIPPIM model, compared to previous FPMD simulations
\cite{giacomazzi2005a,giacomazzi2006a} and to experimental results
\cite{galeener1983a,kamitsos1996a,teredesai2005a}. All the frequencies are given in
cm$^{-1}$.}
\begin{tabular}{|c|ccc|}
\hline
   DIPPIM      & 300 & 601 & 894      \\
   FPMD\cite{giacomazzi2005a}      & 244 & 498 & 787     \\
   Exp. 1\cite{galeener1983a}      & 280 & 567 & 858    \\
   Exp. 2\cite{kamitsos1996a}      & 315 & 585 & 915     \\
   Exp. 3\cite{teredesai2005a}     &  & 560 & 870     \\
\hline
\end{tabular}
\end{table}

\begin{figure}[htbp]
\begin{center}
\includegraphics[width=9cm]{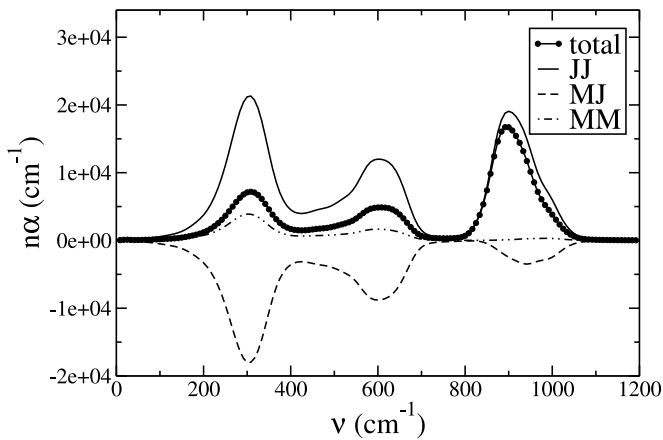}
\end{center}
\caption{ Decomposition of the total IR absorption spectrum in its three component parts. }
\label{decompo}
\end{figure}

The importance of polarization effects in determining the relative intensities of
the bands can be demonstrated by separating the various contributions to the
absorption spectrum. In the case of glassy silica \cite{wilson1996a} and beryllium
fluoride \cite{heaton2006a}, it was observed that the interference between the
induced dipoles and permanent charge contributions to the total polarization,
contained in the $\langle {\bf M}\cdot{\bf J}\rangle_\nu$ cross term, is
responsible for the changes of the relative intensities. Figure \ref{decompo} shows
the decomposition of the absorption spectrum in the case of glassy GeO$_2$. The two main
contributions are the charge fluctuation and the cross term; both show  bands at
the same intensities. Note that the cross-term strongly reduces the intensity of
the two low frequency bands relative to that which would be obtained from the
charge fluctuations alone, whereas for the high-frequency band the cancellation is
much weaker. For the high frequency band the relationship of the charge-charge and
cross-terms is different to that found previously for SiO$_2$ and
BeF$_2$\cite{wilson1996a,heaton2006a}. In the latter cases, the charge-charge and
cross-terms had the same sign, so that the net band intensity was slightly larger
than that predicted by the charge-charge term alone. This difference in the
behaviour of the calculated spectra might arise from the inclusion of an
anion-anion damping term in the polarization part of the interaction potential
here, which was not the case for the previous studies. 

\begin{figure}[htbp]
\begin{center}
\includegraphics[width=9cm]{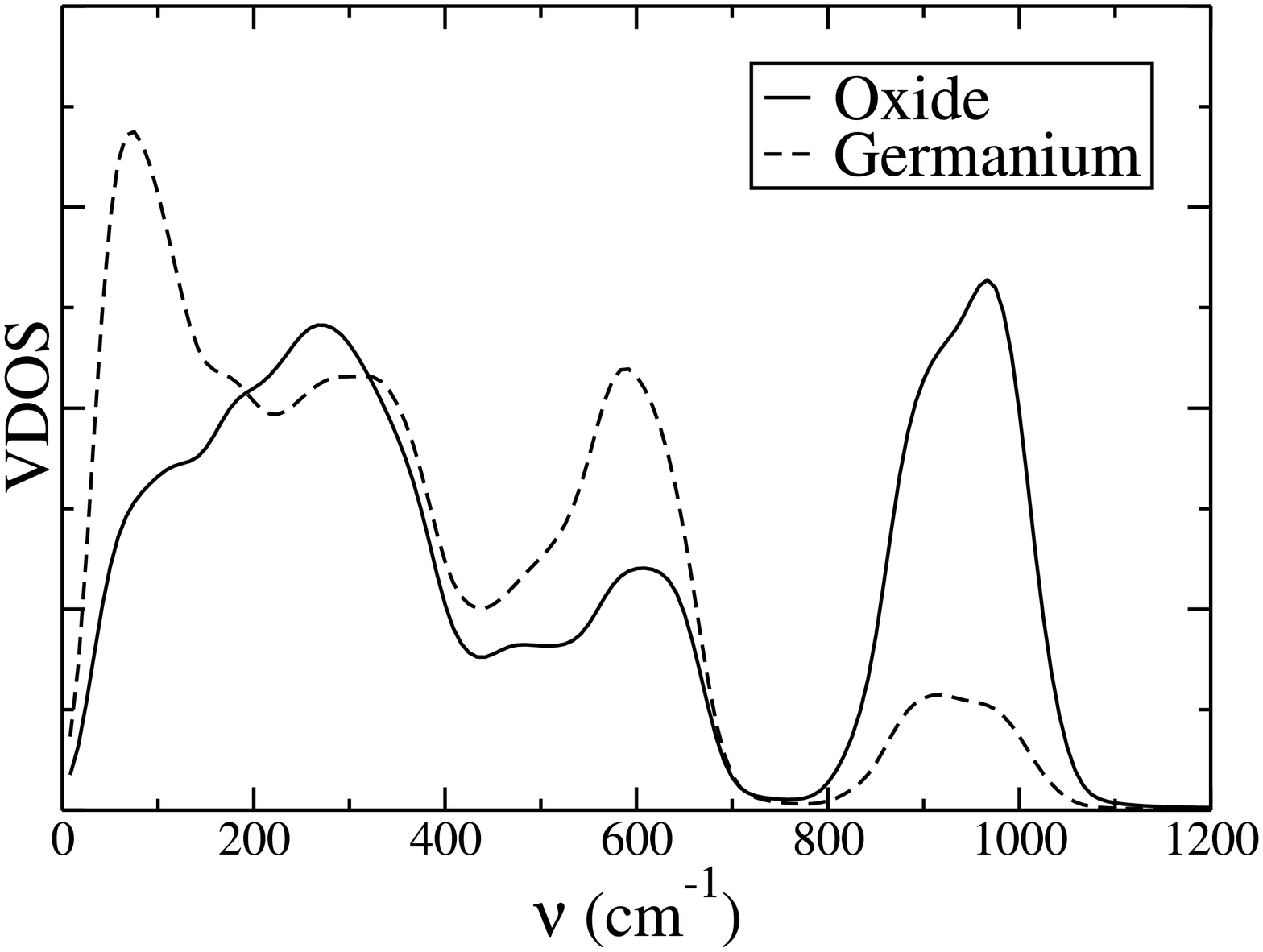}
\includegraphics[width=9cm]{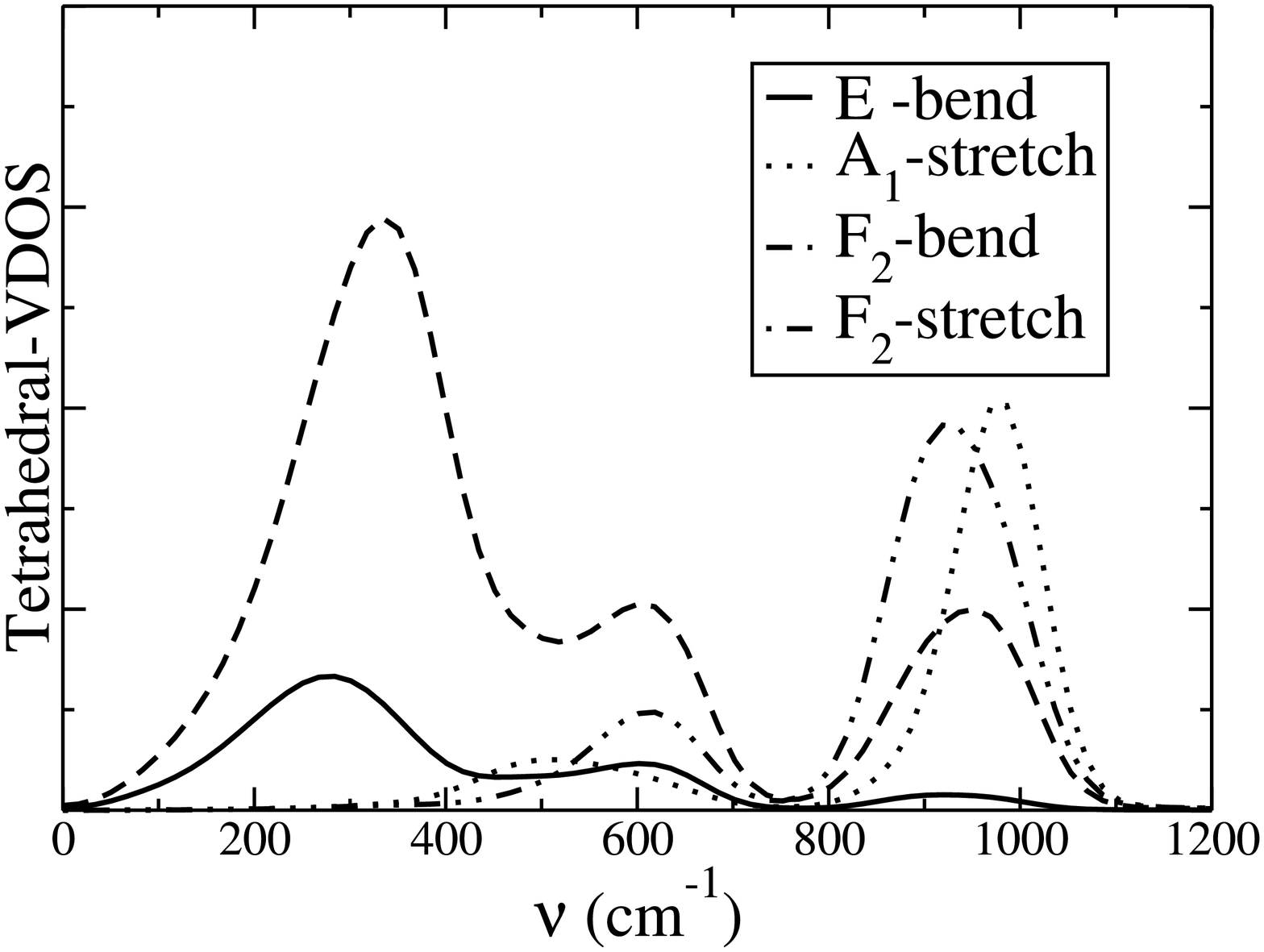}
\end{center}
\caption{ The vibrational densities of states, for the DIPPIM, of both type of ions (top panel), and their projections in terms of the $A$ (symmetric stretch), $F_2$ (asymmetric stretch), $E$ (bend), and $F_2$ (bend) modes of the Ge0$_4$ tetrahedra (bottom panel).}
\label{tdos}
\end{figure}

We can now attempt to associate these  IR frequencies with the vibrational
modes of the system. The upper part of figure \ref{tdos} shows the vibrational density of states (VDOS) of each ion. Some of these features can be associated to the vibrations of the GeO$_4$ tetrahedra by comparing them with the tetrahedral-VDOS as obtained from the 
correlation functions of the velocities of various local tetrahedral symmetry
coordinates. The latter would correspond to the velocities of GeO$_4$ normal modes,
if the tetrahedra were isolated ({\it i.e.} not mechanically linked in a network).
For a tetrahedral molecule, vibrational normal coordinates $\nu_1$ ($A_1$,
symmetric stretch), $\nu_2$ ($E$, bend), $\nu_3$ ($F_2$, asymmetric stretch) and
$\nu_4$ ($F_2$, bend) are expected \cite{pavlatou1997a}. We may obtain velocities
associated with the symmetry coordinates of each GeO$_4$ unit. For example, for the
symmetric stretching motion
\begin{equation}
v^A_i=\sum_{i\alpha=1-4}v^\parallel_{i\alpha}
\end{equation}
where $i$ labels the Ge of a tetrahedral complex, and $i\alpha$  the four oxide
anions at the vertices. $v^\parallel_{i\alpha}$ is the projection of the relative
velocity of $i\alpha$ along the $i\rightarrow i\alpha$ bond, {\it i.e.}
\begin{equation}
v^\parallel_{i\alpha}=({\bf v}_{i\alpha}-{\bf v}_i)\cdot({\bf r}_{i\alpha}-{\bf r}_i).
\end{equation}
Similar expressions may be written down for the  velocities of the other symmetry
coordinates \cite{pavlatou1997a}. The corresponding DOS is then obtained by a
Fourier transform of the corresponding velocity autocorrelation function:
\begin{equation}
{\rm DOS}_A(\nu)=\Re \int_0^{\infty}~ e^{2\pi\imath \nu t}\langle v^A_i(t)\cdot
v^A_i(0)\rangle {\rm d}t.
\end{equation}

Were the network to vibrate as a collection of  isolated tetrahedra, the spectra of
each tetrahedral-VDOS would exhibit a single peak corresponding to the characteristic frequency
of the corresponding normal mode of vibration. The four tetrahedral-VDOS are shown on the lower part of figure
\ref{tdos}, it is clear that it is not the case here. Each spectra consists in the
superposition of several bands, which shows the existence of some important
coupling between the symmetry coordinates. It is therefore difficult to assign the
IR features to the different modes. In particular, all of the three bands observed
in the IR spectrum appear in the tetrahedral-VDOS of the bending $E$ and $F_2$ modes. The other
$F_2$ modes, which is associated to some kind of stretching of the tetrahedra, only
exhibits the medium and high frequency modes. Finally, the $A_1$ stretching
spectrum exhibits two bands at frequencies centered on the values of 510 and
970~cm$^{-1}$, which do not appear on the IR spectrum. Such a mode of this symmetry
is indeed not expected to be IR active for an isolated tetrahedral molecule, since
it corresponds to the symmetric breathing mode. Coming back to the total VDOS of each ion, 
the low frequency vibrations are not associated to any of the tetrahedral modes. In fact these 
are due to some  slow collective bending of two tetrahedra linked by a bridging oxide anion \cite{trachenko2002a}, 
so that they would not appear in the DOS of a single tetrahedron.

\begin{figure}[htbp]
\begin{center}
\includegraphics[width=9cm]{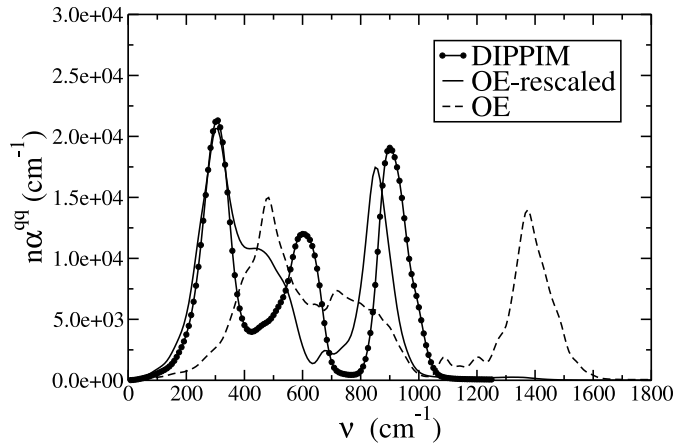}
\end{center}
\caption{ Comparison of the charge-charge  current contributions to the
infrared spectrum obtained with the three different potentials (the two OE curves
were rescaled with a factor $q_{\rm O}^2/4$, where $q_{\rm O}$ is the partial
charge carried by the oxide ion in the corresponding potential. }
\label{epsilonqq}
\end{figure}

The vibrational properties of the system as obtained  with the original and
rescaled OE interaction potentials have also been computed. As these potentials do
not include any polarization effect, it is only possible to compare the
charge-charge current contributions to the IR absorption spectrum with the one
obtained with the DIPPIM potential. The obtained spectra are given on figure
\ref{epsilonqq}. Unsurprisingly, the original OE potential seem to fail completely
in describing the vibrational properties of GeO$_2$; this is the reason why the
rescaled OE potential had to be developed. For that potential, the agreement is
better, in particular for both the small and high frequency bands. Concerning the
medium frequency band, observed at around 604~cm$^{-1}$ for the DIPPIM potential,
it does not appear for the rescaled OE potential. Instead there appears a shoulder
to the small frequency band, for frequencies ranging from 400 to 500~cm$^{-1}$. As
these bands were associated to the $F_2$ stretching and bending modes on the
tetrahedral-VDOS, this means that this modes are not correctly depicted
by the rescaled OE potential.

\section{Conclusion}

In conclusion, the obtained potential seem  to reproduce all the studied properties
(structural, dynamical and vibrational) to a high degree of precision. Since there
was no reference at all to any experimental data in the parameterization of this
potential, this represents a strong test of the model reliability. The study of
pressurized germania is among our top priorities for future work. The potential is
of a similar form to that introduced to describe silicates \cite{jahn2007a}; since
these potentials have proved to be transferable, we will be able, in the future, to
focus on GeO$_2$-SiO$_2$ mixtures as well.  These mixtures, as already mentioned in
the introduction,  attract great interest as they are widely used in optical fibers
and waveguides.

\section*{Acknowledgements}
We thank Philip Salmon for providing his diffraction data. Calculations were
carried out at the EaStCHEM  computing facility (http://www.eastchem.ac.uk/rcf). DM
thanks the EPSRC and the CMPC for his PhD funding. This work was carried out under
the HPC-EUROPA++ project (project number: 211437), with the support of the European
Community - Research Infrastructure Action of the FP7 "Coordination and support
action" Programme.


\end{document}